\documentclass[12pt]{article}

\usepackage{amssymb,amsmath,amsfonts,eurosym,geometry,ulem,graphicx,caption,color,setspace,sectsty,comment,footmisc,caption,pdflscape,subfigure,array,hyperref}
\usepackage[graphicx]{realboxes}
\usepackage{booktabs}
\usepackage{cleveref}
\newcolumntype{L}[1]{>{\raggedright\let\newline\\arraybackslash\hspace{0pt}}m{#1}}
\newcolumntype{C}[1]{>{\centering\let\newline\\arraybackslash\hspace{0pt}}m{#1}}
\newcolumntype{R}[1]{>{\raggedleft\let\newline\\arraybackslash\hspace{0pt}}m{#1}}

\begin{document}


\title{Foundations of Cyber Resilience: The Confluence of Game, Control, and Learning Theories}

 \author{Quanyan Zhu}

\maketitle

\abstract{
Cyber resilience is a complementary concept to cybersecurity, focusing on the preparation, response, and recovery from cyber threats that are challenging to prevent. Organizations increasingly face such threats in an evolving cyber threat landscape. Understanding and establishing foundations for cyber resilience provide a quantitative and systematic approach to cyber risk assessment,  mitigation policy evaluation, and risk-informed defense design. A systems-scientific view toward cyber risks provides holistic and system-level solutions. This chapter starts with a systemic view toward cyber risks and presents the confluence of game theory, control theory, and learning theories, which are three major pillars for the design of cyber resilience mechanisms to counteract increasingly sophisticated and evolving threats in our networks and organizations. Game and control theoretic methods provide a set of modeling frameworks to capture the strategic and dynamic interactions between defenders and attackers. Control and learning frameworks together provide a feedback-driven mechanism that enables autonomous and adaptive responses to threats. Game and learning frameworks offer a data-driven approach to proactively reason about adversarial behaviors and resilient strategies. The confluence of the three lays the theoretical foundations for the analysis and design of cyber resilience. This chapter presents various theoretical paradigms, including dynamic asymmetric games, moving horizon control, conjectural learning, and meta-learning, as recent advances at the intersection. This chapter concludes with future directions and discussions of the role of neurosymbolic learning and the synergy between foundation models and game models in cyber resilience.
 }

\section{Introduction to Cyber Resilience}

 Cyber resilience of a network or organization refers to its ability to prepare for, respond to, and recover from cyber threats or incidents \cite{ganin2016operational,zhao2022multi,rieger2012agent,rieger2009resilient}. It represents a departure from classical cybersecurity measures, including cryptography, firewalls, and intrusion detection systems, which aim to prevent attackers from infiltrating or causing harm to the network. Instead, cyber resilience focuses on the capacity to respond to and recover from threats. Cyber resilience and cybersecurity are complementary to each other. Despite efforts and investments in cybersecurity for protection, perfect security cannot be guaranteed. There will always be threats that cannot be fully mitigated. These threats may include zero-day attacks, which are unknown to the network, or black swan events that are costly to secure against. Therefore, with limited resources and budgets, networks must decide how to protect themselves from threats to ensure the continuity of their applications and missions. 

As illustrated in Figure \ref{distinction}, resilience provides a complementary approach \cite{zhu2015game,zhu2020cross,10.1145/3549073}. When threats are difficult to prevent successfully, there is a need to find ways to respond to them after they have gained initial access to the network, deter further infiltration by attackers, and recover from any damage incurred. Additionally, it is important to recognize that the success of security mechanisms is probabilistic. A higher probability of success will require a greater allocation of resources. Despite the careful deployment of cybersecurity measures, there will still be a probability, albeit small, for attackers to succeed in bypassing cryptographic protocols and firewalls to gain access to the network.

\begin{figure}[ht]
\begin{center}
    \includegraphics[scale=0.7]{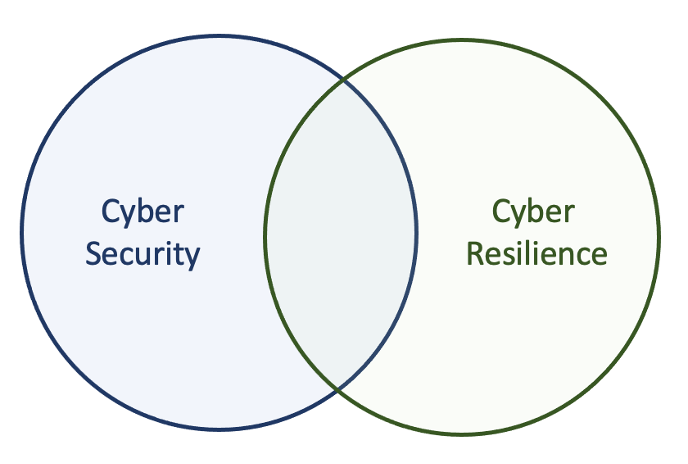}
    \caption{Cybersecurity and cyber resilience focus on different sets of issues. Cybersecurity primarily concentrates on safeguarding assets from unauthorized access, attacks, and damages. On the other hand, cyber resilience aims to both deter attackers and recover the network from the initial success of attacks. Both strive to enhance the confidentiality, integrity, and availability (CIA) of information and resources. Cyber resilience specifically aims to mitigate the impact on the CIA triad despite the initial compromise. Despite their differences, there are overlaps between the two in terms of the techniques used to provide security and resilience. For instance, detection plays a critical role in both security and resilience efforts. }
\end{center}
\end{figure}\label{distinction}

 The tradeoffs between cyber resilience and cybersecurity stem from economic and technological considerations. As illustrated in Figure \ref{vulnerability}, common threats that occur frequently or are easily exploited by attackers often require defensive security mechanisms to protect the network with a high probability of success. Conversely, sophisticated threats that demand advanced skill sets and occur less frequently can be mitigated through defensive resilience. The decision on resource allocation across the entire threat landscape depends on factors such as budget, application domains, and available technology. For instance, mission-critical applications like nuclear power plants or oil and gas manufacturing plants require investment not only in designing security mechanisms but also in crafting resilience mechanisms to counter highly sophisticated attackers and mitigate the exploitation of unknown vulnerabilities.

The decision regarding which threats to defend against is not solely limited by budgetary constraints; it also involves human resources, such as the time and effort required for defense, as well as the availability of technologies that can facilitate or enable such defense. For instance, defending against Advanced Persistent Threat (APT) attacks requires a coordinated effort from various aspects of missions, including human, cyber, and physical assets, to ensure assurance. This defense may require AI technologies, which may still be in their infancy, particularly those that can be employed to mitigate the exploitation of human cognitive vulnerabilities leading to successful social engineering attacks.

Within the set of exploitable vulnerabilities, we categorize them into two types of risks: avertable risk and elastic risk. Elastic risks are those triggered by events that can only be mitigated elastically through resilience strategies. Conversely, avertable risks are triggered by events that can be prevented with high probability using security mechanisms.

\begin{figure}[ht]
\begin{center}
    \includegraphics[scale=0.7]{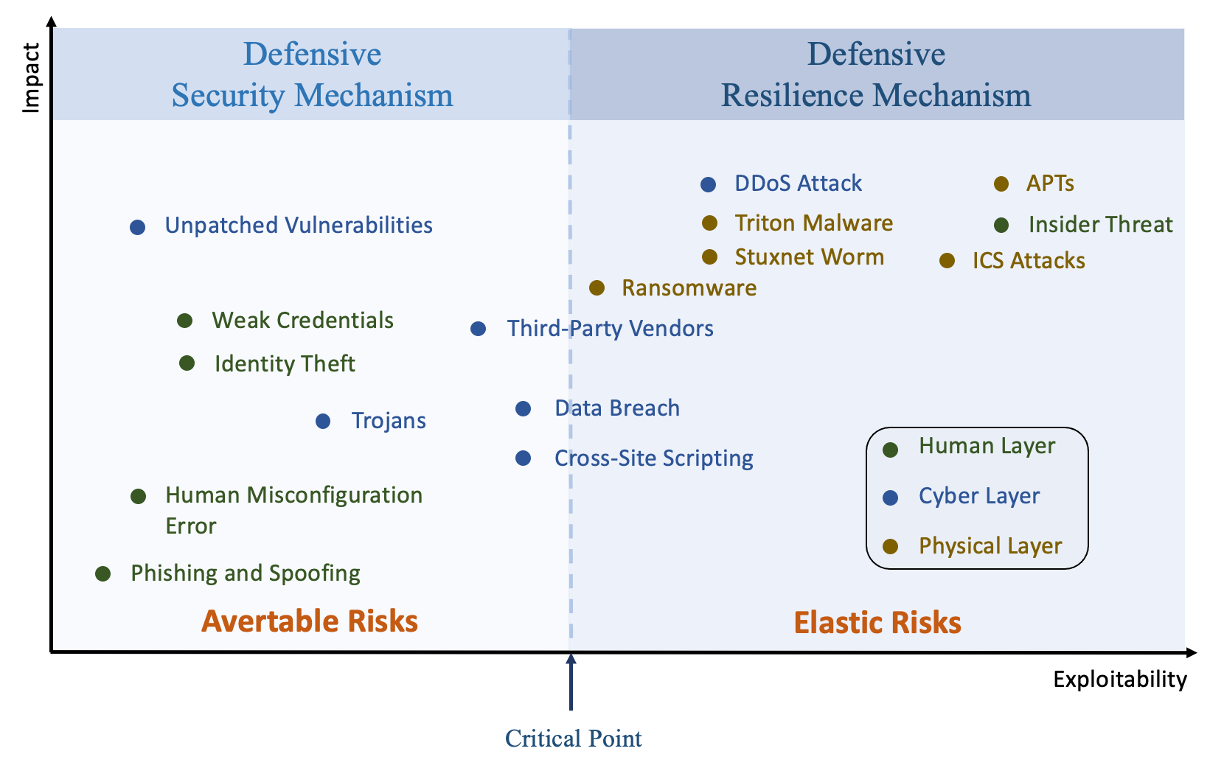}
    \caption{The vulnerabilities and their associated threats on the attack surface are represented in two dimensions: their exploitability and their impact. They span across cyber, physical, and human layers of networks or organizations. These vulnerabilities are categorized into two types of risks. One type is avertable risks, which are managed by defensive security mechanisms, while the other type is elastic risks, which are addressed through resilience mechanisms. }
\end{center}
\end{figure}\label{vulnerability}

\section{Cyber Risks and Resilience Mechanisms}
The first step toward understanding cyber resilience is to comprehend and quantify cyber risks. Just like all other risks, cyber risks are characterized by three primary components. First and foremost, cyber risk stems from the existence of vulnerabilities within the network. These vulnerabilities can manifest as software bugs, human biases, interface errors, system integration failures, inadequate preparedness, or hardware fragility. They serve as sources of susceptibility or exposure for both the network and the organization to identified threats.

These vulnerabilities create risks because they represent opportunities for threats to exploit them. Hence, the threat constitutes the second component of the risk. A threat refers to an event or circumstance created by an adversary to exploit vulnerabilities within the system, triggering further events that lead to harm, loss, or damage. For instance, an attacker may craft a phishing email to exploit the cognitive vulnerability of a user, leading to the disclosure of passwords or access credentials to the attacker, or the unwitting download of malware to spread within the network. Alternatively, an attacker may exploit a buffer overflow bug in software to execute malicious code.

The intersection of vulnerabilities and threats forms what we term the attack surface. This represents the collection of vulnerabilities and entry points within the network or organization that attackers can potentially leverage to gain unauthorized access, disrupt operations, steal data, or carry out other malicious activities.

When an adversarial event exploits a vulnerability within the attack surface, consequences ensue. Consequences represent the outcomes of the interaction between threats and vulnerabilities. They can encompass performance degradation, disruptions to operations, financial losses, reputational damage, legal liabilities, and other adverse effects. 

The vulnerabilities, threats, and consequences constitute the landscape of cyber risks, all of which can evolve over time. As illustrated in Figure \ref{risk}, the set of vulnerabilities at time $t$ is denoted by $V_t$, while the set of threats is denoted by $T_t$. One immediate consequence of the exploitation by an adversary event, $E_t$, is the change of the network system state, $X_t$, which subsequently influences the consequences for the network, denoted by $Y_t$.

A vulnerability (3) is one that resides on the attack surface. For each vulnerability on the surface, there exists an event that leads to exploitation. However, a vulnerability, as exemplified in (2), is a threat that does not result in an event capable of exploiting a vulnerability within the network. This could be due to the network being well protected against that threat or the threat being irrelevant to the system under focus. A vulnerability (2) is one that cannot be exploited by a threat, either because the adversary is unaware of it or the vulnerability is difficult to exploit.

A cyber resilience mechanism is one that supports the network system in dynamically reducing its risk over time through three distinct types of cyber resilience mechanisms. One such mechanism is the proactive approach, which focuses on directly changing the attack surface to reduce the risk. Proactive measures may involve creating secure network policies, such as network segmentation, patch management, and zero trust access control to minimize potential threat vectors.  For example, network segmentation involves dividing a network into smaller, isolated segments, organizations can limit the scope of potential attacks. Network segmentation allows for stricter access controls and segmentation of sensitive resources, preventing lateral movement by attackers within the network.  Regularly updating and patching software and systems helps address known vulnerabilities and weaknesses. Patch management programs ensure that critical security patches are applied promptly to reduce the risk of exploitation by attackers. Zero-trust access control verifies the identity of users and devices attempting to access resources, regardless of their location or network connection. This involves implementing strong authentication methods such as multi-factor authentication (MFA) and biometric authentication to ensure that only authorized individuals gain access. By proactively implementing these policies, the set of vulnerabilities $V_t$ or threats $T_t$ can be reduced, subsequently shrinking the attack surface.

Another one is responsive resilience mechanisms that  involve continuous monitoring and assessment of the threat landscape to identify emerging risks and vulnerabilities and dynamic defense strategies that evolve in real-time based on contextual factors and situational awareness. By leveraging predictive analytics and threat intelligence, organizations can anticipate potential attack vectors and vulnerabilities, allowing them to strengthen their defenses before adversaries can exploit them. The dynamic adaptability enables organizations to quickly adjust their security posture in response to evolving threats and changing environmental conditions. Leveraging machine learning and control algorithms, these adaptive defenses can autonomously analyze incoming threats, assess their impact, and dynamically allocate resources to mitigate them effectively. One notable example is adaptive cyber deception \cite{jajodia2016cyber,pawlick2018modeling,zhang2020game,pawlick2017proactive}, aimed at thwarting attackers from achieving their objectives by deploying decoys, such as honeypots or other traps, to divert their attention from critical assets. The effectiveness of cyber deception schemes relies on their adaptability, achieved through continuous learning of attackers' objectives and tactics. By dynamically configuring and deploying deception techniques based on this acquired knowledge, organizations can enhance their ability to mislead and deter adversaries effectively.  Moving target defense is another example that can also be tailored to counteract attackers' evolving tactics and behaviors. Moving target defense such as network address randomization, where IP addresses and network configurations are frequently shuffled to prevent attackers from mapping out the network topology and identifying vulnerable entry points. System administrators can employ dynamic service migration, automatically moving critical services and applications between different hosts or cloud instances to evade detection and mitigate the impact of targeted attacks.

The third mechanism is retrospective, which focuses on restoring the system after an attacker has successfully compromised it. One immediate retrospective mechanism is cleaning up the network following a shutdown. Consider a manufacturing plant that falls victim to a sophisticated ransomware attack. As a result of the attack, critical systems controlling production lines and machinery are rendered inoperable. To contain the spread of the ransomware and prevent further damage, the plant is forced to initiate a complete shutdown of its operations. This entails halting production, isolating infected systems, and disconnecting the compromised network from external connections.  The restoration process involves restoring encrypted data from backups, rebuilding compromised systems, and implementing enhanced security measures to prevent future incidents. The recovery process incurs a significant investment of time, resources, and financial costs. 

Another retrospective mechanism is post-incident analysis and remediation efforts aimed at understanding the root cause of the breach and implementing measures to prevent similar incidents in the future. This involves conducting thorough forensic investigations to identify the entry point of the attack, the techniques used by the adversary, and any vulnerabilities exploited. By analyzing the attack vector and identifying weaknesses in the security infrastructure, organizations can implement targeted security enhancements and patches to strengthen their defenses against future threats. Cyber insurance is another example of a retrospective mechanism by organizations to mitigate the financial fallout of a cybersecurity incident \cite{liu2023cyber,zhu2018cyber}. In the aftermath of a data breach or network intrusion, businesses may face a multitude of costs, including forensic investigations, legal fees, regulatory fines, and loss of revenue. Cyber insurance policies provide a financial safety net by offering coverage for these expenses and compensating organizations for the losses incurred as a result of the incident. By having a cyber insurance policy in place, organizations can alleviate the financial burden of cybersecurity incidents and safeguard their long-term viability.  

While retrospective mechanisms are often a last resort, they become necessary when facing a formidable adversary. Proper preparation and design are essential to minimize recovery time and mitigate financial and technical risks to the network.

\begin{figure}[!h]
\begin{center}
    \includegraphics[scale=0.7]{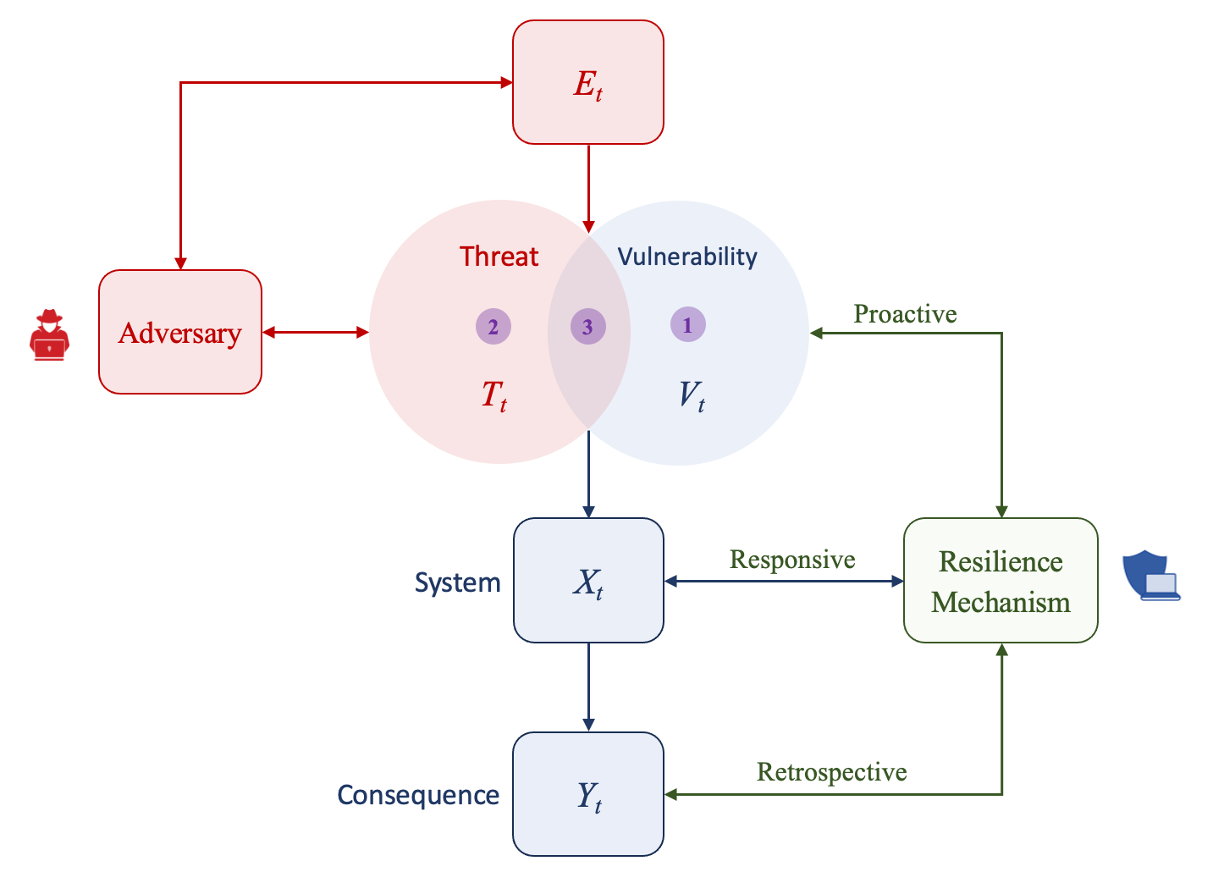}
    \caption{Cyber risk at time $t$ arises from vulnerabilities within the system itself, denoted by the set $V_t$, and potential threats posed by adversaries $T_t$. The intersection of these threats and vulnerabilities forms the attack surface of the network. An adversary can launch an event, $E_t$, to exploit a vulnerability on the attack surface, resulting in changes to the system states. The consequences are expressed as outcomes of these changes in system state. The goal of resilience mechanisms is to mitigate the impact of such attacks. There are three types of resilience mechanisms: proactive, responsive, and retrospective. Proactive mechanisms aim to reduce the attack surface and create a more resilient network that is harder for attackers to exploit. Responsive mechanisms involve equipping the network with adaptive and automated responses to attack behaviors in real-time. Retrospective mechanisms focus on reducing the impact after the consequences of the attack, $Y_t$ , are realized and observed. They can involve restoring the system to its previous state or recovering financially from losses incurred.}
\end{center}
\end{figure}\label{risk}

\section{Network and Dynamic Effects on Cyber Risks}

A large network can consist of multiple interconnected subnetworks. The risk within one subnetwork can be influenced by others, as they are interconnected. A threat to one subnetwork can lead to consequences within that subnetwork and may also impact other connected subnetworks. As illustrated in Figure \ref{influence}, the states of other subnetworks are aggregated as $\Sigma_t$, and the well-being of these other subnetworks can affect the state of the subnetwork $X_t$. The interdependencies among subnetworks pose a significant challenge in managing cyber risks. Risk in one subnetwork can propagate to other subnetworks and vice versa. These interdependencies can arise from three sources.  Firstly, threats can exploit the connectivity between subnetworks, making it more convenient for attackers to deploy threats. For instance, attackers may compromise neighboring subnetworks before targeting the intended one. In this way, the threat space $T_t$ is expanded for the targeted network as threats to connected networks also affect its security. Secondly, vulnerabilities in the connections with neighboring subnetworks bring in new vulnerabilities of the subnetwork itself. Additionally, vulnerabilities in the nodes connected to it become potential threats that the subnetwork must address. In this way, the vulnerability space $V_t$ is expanded for the targeted network. Thirdly, interdependencies can arise through consequences. Risks can propagate through state influence, where the state of one subsystem influences others through physical dynamics. In this way, the consequence $X_t$ or $Y_t$ is directly affected.

Such interdependencies have been vastly investigated in critical infrastructure networks \cite{chen2019game,huang2018distributed,zimmerman2017conceptual,huang2017large,zimmerman2016promoting} and communication networks \cite{farooq2018multi,farooq2018adaptive,farooq2019modeling,kieras2022iot}. Defending against threats originating from within the network itself is challenging, and preparing for risks from connected networks is even more complex. Subnetworks may face threats from other networks for which they are not directly prepared. Therefore, resilience is critical in addressing these challenges, particularly due to the unknown risks that propagate through network connections.

\begin{figure}[!h]
\begin{center}
    \includegraphics[scale=0.7]{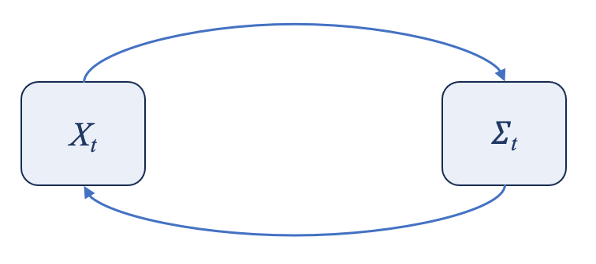}
    \caption{In a networked environment, the system states are influenced not only by its own risk factors but also by the system states of other nodes connected to it. $\Sigma_t$ denotes the aggregate state of other nodes, while $X_t$ represents the state of the system itself. The risk of a network system arises from not only local internal risks but also interdependent risks due to these connections. Studies on interdependent risks have been explored in the literature. One way to understand such risks is through game-theoretic methods, which have been studied for systemic risks \cite{acharya2017measuring} and holonic risks \cite{liu2022herd,van1998reference}.}
\end{center}
\end{figure}\label{influence}

Resilience mechanisms must be dynamic in nature as they operate within dynamic environments. Firstly, threats evolve over time. Unlike natural threats, adversaries can discover new vulnerabilities and acquire new knowledge to expand the attack surface. Figure \ref{threat1} depicts the security landscape at time 1, denoted by threat set T1 and vulnerability set $V_1$. Figure \ref{threat2} also illustrates the security landscape at time 2, with the corresponding threat set T2 and vulnerability set $V_2$. It is evident that the threat set expands as adversaries learn and enhance their skill sets over time. Conversely, the defender's vulnerability set increases as the network introduces new vulnerabilities due to software and patch updates, as well as changes in connectivity. Consequently, the attack surface dynamically grows from time 1 to time 2. Such growth can be strategically directed by the actions taken by both parties. The attacker may choose actions to expand its set of attack vectors by probing the network, while defenders can take actions to increase or decrease their vulnerabilities through the implementation of defense mechanisms over time. Therefore, resilience mechanisms must adapt over time to address the evolving landscape of threats and the dynamics of the system itself. This adaptation encompasses not only responsive mechanisms but also proactive and retrospective ones. One distinction is that they may evolve at different time scales.

Resilience mechanisms must also consider incomplete knowledge and uncertainties associated with threats and vulnerabilities. The complete set of vulnerabilities, as depicted in Figure \ref{threat1}, is not always known to the agents. Agents possess varying degrees of knowledge regarding vulnerabilities. One example is zero-day vulnerabilities, where the network defends itself without awareness of their existence in the software or hardware. If an attacker exploits zero-day vulnerabilities, they can create unexpected events that the defender did not anticipate, leaving the network defenseless. In such cases, resilience mechanisms become the last resort to mitigate such exploitation. The asymmetric information or knowledge between the attacker and defender adds another layer of complexity that resilience mechanisms must address.

In addition to knowledge asymmetry, uncertainties exist in the network environment such as user behaviors, traffic patterns, and topology changes. For example, in a battlefield communication network \cite{farooq2017secure,farooq2018secure,farooq2021resource},  uncertainties in terrain features, such as hills, forests, and urban structures, can lead to signal interference, propagation losses, and unpredictable coverage areas. In addition, uncertainties in military unit movements, deployments, and formations can disrupt communication links. These uncertainties can affect the probability of threat event success and consequently impact the state of the network. Resilience mechanisms require tools capable of handling such uncertainties. While investing in defense can reduce the probability of certain events succeeding or the attacker advancing to the next level of attacks, the risk may still persist, and resilience is the key approach to mitigate it.

\begin{figure}[!h]
\begin{center}
    \includegraphics[scale=0.7]{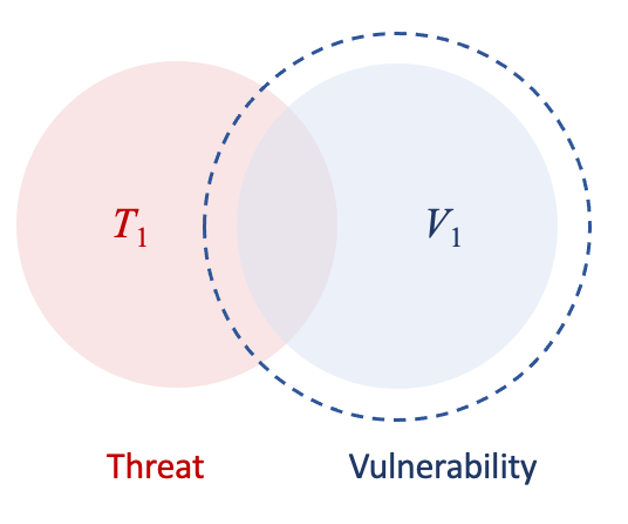}
    \caption{The vulnerability set $V_1$ can be significantly larger, unbeknownst to the network defender, due to incomplete knowledge of vulnerabilities. This implies that the attack surface can be much broader than what is currently known. Attackers may exploit vulnerabilities on the attack surface that are unknown to defenders. These types of vulnerabilities must be addressed through cyber resilience strategies.}
\end{center}
\end{figure}\label{threat1}

\begin{figure}[!h]
\begin{center}
    \includegraphics[scale=0.55]{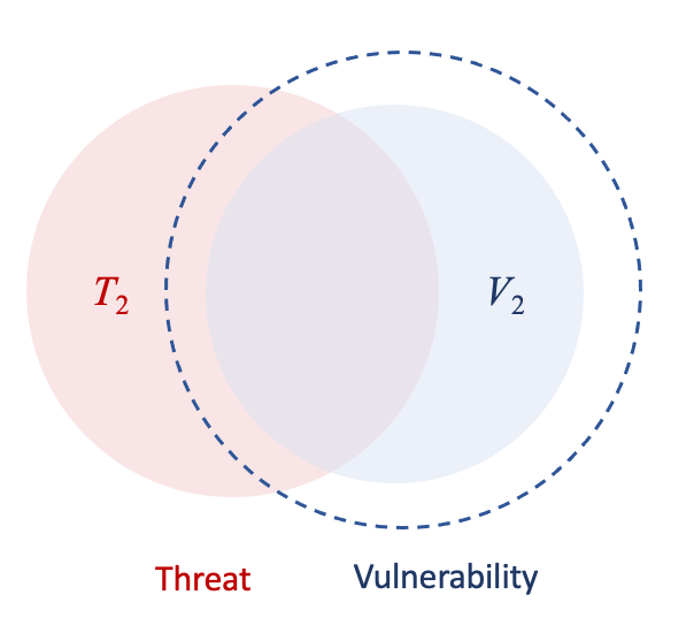}
    \caption{At stage 2, the network may acquire new vulnerabilities through the installation of new software, patching, or the addition of new interfaces. Consequently, the attack surface expands accordingly. The security landscape evolves from stage 1 to stage 2. There is a need for cyber resilience to ready the network for newly introduced vulnerabilities. Preparing to protect against these vulnerabilities in real-time is challenging due to the dynamic nature of network changes.}
\end{center}
\end{figure}\label{threat2}

To design dynamic mechanisms, control and learning tools are indispensable. This entails monitoring systems and gathering information about adversaries. Based on this data, reasoning frameworks are developed to respond to the system, as illustrated in Figure \ref{dynamics}. The sequential events initiated by the attacker and the resilience policies implemented by the defender both influence the state of the network. The objective of resilience mechanism design is to regulate the state of the network to achieve acceptable performance in its evolution.

\begin{figure}[ht]
\begin{center}
    \includegraphics[scale=0.7]{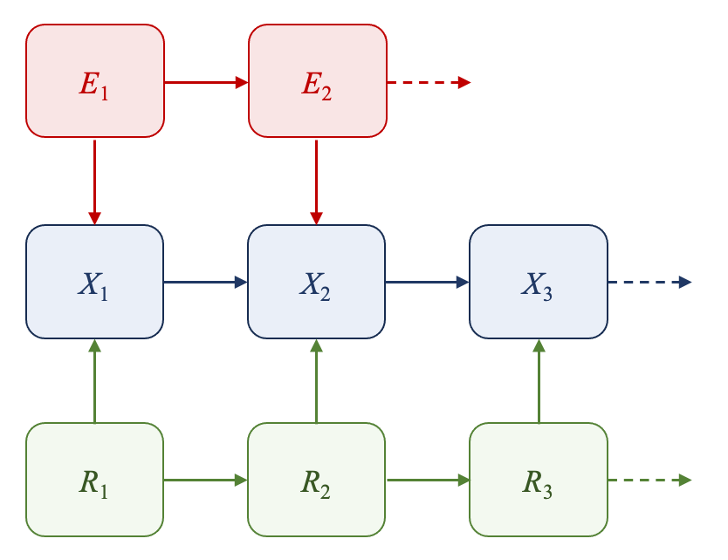}
    \caption{Sequential interactions between the adversarial exploitation events $E_1, E_2$ and the actions taken by the resilience mechanisms $R_1, R_2, R_3$. They affect how the system states evolve over time $X_1, X_2, X_3$.}
\end{center}
\end{figure}\label{dynamics}

\section{Cyber Resilience and Cyber Risk Metrics}

Before designing, it is essential to establish metrics for quantifying performance and design criteria. Figure \ref{metric} illustrates the evolution of network performance over time, driven by sequential interactions between the attacker and the defender, as depicted in Figure \ref{dynamics}. Performance over time is represented by the measurable performance $Y_t$ across various stages. The evolutionary process is divided into three phases: proactive, responsive, and retrospective. At $t_1$, the proactive mechanism is implemented. Subsequently, at $t_2$, an adversary exploits vulnerabilities in the attack surface, leading to performance degradation at $t_3$. Responsive defensive measures at $t_4$ restore the network's performance from level $M$ to level $D$. The retrospective mechanism engages at $t_5$ to address post-incident damage and restore performance.

Numerous metrics can be proposed to quantify resilience, which can be classified into three categories. The first category is performance-based metrics. These metrics require the monitoring of system states and provide accurate measures of resilience when credible information about system states is available. For example, the total losses incurred between $t_2$ and $t_5$, denoted by $J_E$, reflect the impact of the event initiated by the attacker. Other metrics in this category include $M$ and $D$, which assess the severity of damage and the effectiveness of recovery during the responsive stage. The second category comprises latency-based metrics, which are dependent on timing. These metrics measure the speed or frequency of reaction against risks. For instance, the response time $t_4-t_2$ indicates how quickly recovery measures take effect, while the retrospective time $t_5-t_2$ reflects the time taken for the system to rectify the problem retrospectively. The third category consists of knowledge-based metrics, which are associated with the known area of the attack surface (AS). Resilience in this context depends on the comprehensiveness of the knowledge about the system. For example, the percentage of the system model that can be expressed deterministically.

Resilience metrics offer a means to gauge risk within our network under a specified resilience mechanism. The risk associated with vulnerability $v_E$ comprises two key components: impact and the probability of successful exploitation. With the integration of resilience mechanisms into the network, impact can be quantified by the total loss $J_E$ resulting from the successful exploitation of $v_E$. The probability of success, denoted by $p^t_E$, is influenced by security mechanisms and proactive resilience measures. Risk is expressed as the product $r^t_E = p^t_E \times J^t_E$, and overall risk is evaluated by aggregating $r^t_E$ across all events in the attack surface (AS), which is denoted by

\begin{equation}\label{riskmetric}
r^t=\sum_{E\in AS}\ r^t_E. 
\end{equation}

While this formula does not consider the cumulative effect of events, it serves as a fundamental metric for evaluating cyber risk. It is noteworthy that $p^t_E$ evolves over time and is shaped by the tactics employed by both attackers and defenders. The security and resilience strategies employed by the parties not only impact the probability but also affect the magnitude of impact. Thus, the objective of cyber resilience, considering the implemented security strategies, is to mitigate both overall and long-term risks, as captured by \ref{riskmetric}. A baseline design problem can be finding resilience policies $\{R_t\}$ that minimize $\sum_{t=1}^H r^t$ over the horizon from stage 1 to stage $H$.

It is evident that the problem presents an optimal control challenge, wherein the dynamics are governed by the evolution of cyber states, manipulated by both parties as depicted in Figure \ref{dynamics}. However, solving this problem is inherently difficult. Firstly, it is necessary to incorporate the attacker's behavior into the control problem to more accurately represent the dynamics. To achieve this, adopting a game-theoretic perspective towards this fundamental problem is essential. The evolving interactions over time can be captured through a dynamic game framework involving two players. Secondly, the defender does not possess complete knowledge of the dynamics and the attacker's behaviors. Despite employing game-theoretic models, there persists a challenge in accurately determining the parameters of the underlying game. Thus, learning becomes indispensable in bridging the gap between theoretical game frameworks and practical implementation. Consequently, the framework of cyber resilience and its design is built upon the intertwined principles of control, game theory, and learning, which are essential for assessing, analyzing, and designing for cyber resilience.

\begin{figure}[ht]
\begin{center}
    \includegraphics[scale=0.7]{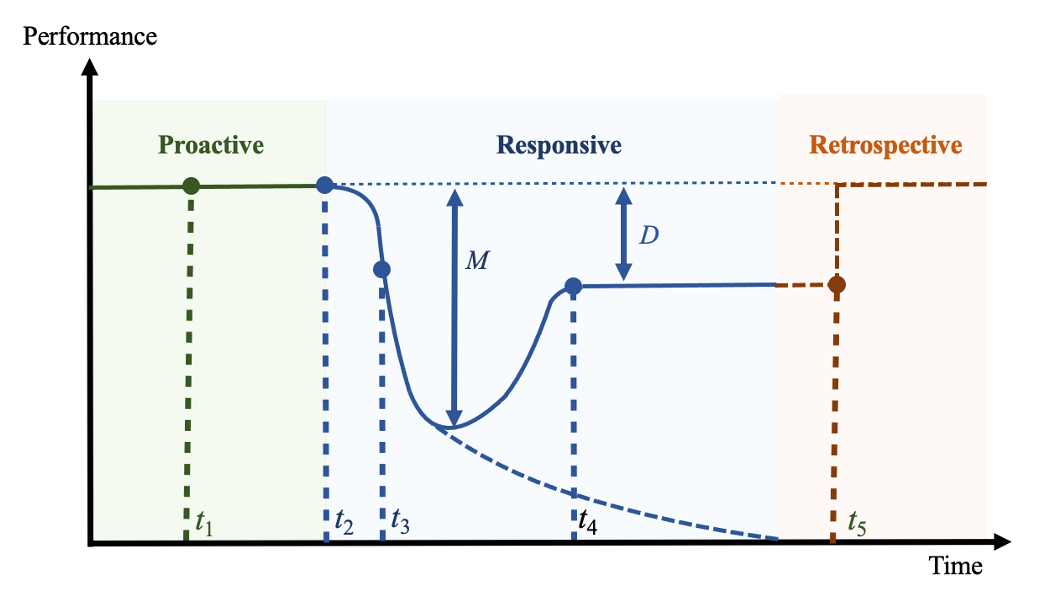}
    \caption{The metrics of cyber resilience in the time domain divide the performance of the network into three stages: proactive, responsive, and retrospective.  In the proactive stage, resilience mechanisms prepare strategies to deter potential attacks.  The responsive stage involves online, adaptive responses to adversarial behaviors. The retrospective stage focuses on resolving problems or minimizing further impact after an incident has occurred and caused damage. $t_1$ represents the time when the preparation mechanism is deployed and implemented. $t_2$ marks the beginning of an event initiated by the adversary to exploit vulnerabilities on the attack surface.  $t_3$ indicates the point when performance degradation begins. $t_4$ is the moment when responsive defense measures facilitate the network's recovery to performance level $D$ after a significant performance drop to level $M$. Without the responsive resilience mechanism, performance degradation would have continued. The retrospective mechanism becomes active at $t_5$, aiming to restore performance after the incident has caused damage.}
\end{center}
\end{figure}\label{metric}

\section{Theoretical Foundations}
Following the discussion above, control theory, game theory, and learning provide theoretical foundations for understanding cyber resilience. Control theory addresses the dynamic nature of threat landscapes and network systems by offering tools to shape the metrics associated with cyber resilience. Game theory naturally captures the adversarial and strategic interactions between defenders and attackers, providing models and tools to analyze and predict the outcomes of these interactions and to design resilience mechanisms and associated strategies to achieve performance objectives despite strategic exploitations and disruptions by attackers. Learning is indispensable for enabling the system to assimilate data, information, and knowledge, allowing the integration of control and game-theoretic algorithms with them. Learning leads to implementable algorithms despite unknowns and uncertainties in the model. In this section, we discuss these foundations separately and argue that their confluence provides theoretical foundations for quantifying, analyzing, predicting, and designing cyber resilience for networks.

\subsection{Control-theoretic foundations}

The formulated control problem provides a baseline for understanding and designing cyber resilience mechanisms. In Figure \ref{riskexample}, we illustrate the evolution of the attack surface over time. The attacker exploits vulnerabilities on this surface, progressing stage by stage towards the targeted asset. A resilience mechanism can eliminate the vulnerability targeted by the attacker in stage 2, thereby deterring their progress. An example is the moving target defense, which alters the attack surface by changing network attributes such as IP addresses, ports, or software configurations. Importantly, this process is dynamic. A successful attack at stage 1 results in a new attack surface at stage 2. To shift the attack surface or eliminate targeted vulnerabilities, defenders must understand the current situation, including the vulnerability set, successful attacker exploits, and the network's new state. Additionally, defenders can analyze the attacker's objectives and intended assets to determine how best to alter the attack surface. However, reconfiguring and changing networks to shift the attack surface incurs costs. Thus, there is a need to find implementable approaches that maintain system performance while minimizing defense costs.

\begin{figure}[ht]
\begin{center}
    \includegraphics[scale=0.7]{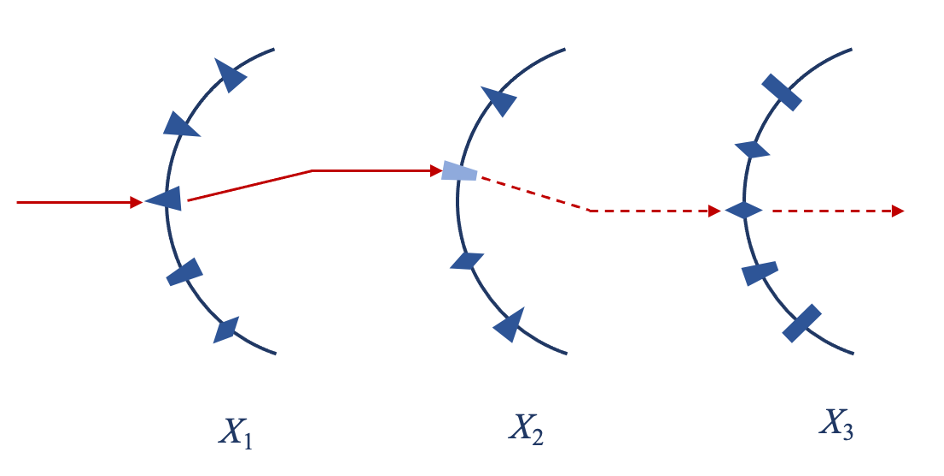}
    \caption{The evolution of the security landscape results in different attack surfaces at various stages. The attacker can exploit the initial vulnerability at stage 1, causing the system state to transition from $X_1$ to $X_2$. This attack can be thwarted by resilience mechanisms if proactive measures eliminate a vulnerability (targeted by the attacker) from the attack surface at stage 2, thereby disrupting the entire attack progression. Reactive mechanisms also enable the system to reconfigure and present a different attack surface (e.g., through moving target defense or cyber deception), deterring the attacker from further progression. Resilience mechanisms, by nature, must consider multi-stage dynamic behavior, with control theory being an appropriate framework for effectively addressing this aspect.}
\end{center}
\end{figure}\label{riskexample}

This approach coincides with the concept of feedback in control theory, where sensing, control, and actuation form the core components. This provides an essential building block for cyber resilience mechanisms. There is a need for situational awareness to comprehend the network's state, reasoning to configure responses subject to system constraints, and triggers to initiate changes. The concept of feedback loop can be further extended to an OODA loop consisting of the stages of ``Observe", ``Orient", ``Decide", and ``Act", \cite{zager2017ooda,bodstrom2018novel}. The ``Observe" stage involves gathering information about the network, including identifying threats, vulnerabilities, and changes in the landscape. Observations are made through various means, such as monitoring network traffic, analyzing system logs, and gathering threat intelligence. The ``Orient" stage, the collected information is analyzed and interpreted to understand the current situation and assess its implications. This includes evaluating the capabilities of adversaries, identifying potential risks, and understanding the context in which decisions need to be made. In the ``Decide" stage,  based on the observations and orientation, decisions are made regarding the appropriate course of action. This stage involves evaluating different options, weighing their potential outcomes, and selecting the most effective response strategy. Finally, in the ``Act" stage,  the decided course of action is implemented. This may involve deploying security measures, executing response plans, or initiating countermeasures to mitigate threats and vulnerabilities.
The control-inspired OODA loop allows for continuous improvement and adaptation, enabling organizations to stay ahead of evolving threats and maintain a resilient posture in the face of cyber challenges.

Besides the concept of feedback, many control tools can become available to solve the problem. For example, the problem can be addressed using moving horizon techniques, which is a control strategy used in dynamic systems to optimize future control actions while considering a finite prediction horizon. The system's current state is observed, and a prediction model is used to forecast the system's behavior over a certain time horizon into the future. Based on this prediction, an optimization problem is formulated to determine the optimal control inputs over this finite horizon that minimizes a certain cost function or satisfies a set of constraints.  In \cite{ge2023gazeta}, moving horizon computation has been used to solve a partially observable Markov decision process, continuously updating and adjusting its trust evaluation based on the latest available information. Additionally, dynamic trust allows for access control decisions over time, creating a zero-trust architecture for 5G networks in the face of evolving threats.

Moreover, \cite{qian2020receding} studies the problem of assessing the effectiveness of a proactive defense-by-detection policy with a network-based moving target defense. A moving horizon planning algorithm proposed in \cite{qian2020receding} is employed to compute the attacker's strategy for security evaluation. This approach involves continually updating the attacker's plan over a finite time horizon based on real-time information and changing conditions. As the planning horizon advances, the algorithm recalculates the optimal strategy, taking into account the latest insights and observations.

\subsection{Game-theoretic foundations}

As cyber risk emerges from strategic interactions between players, resilience mechanisms aiming to mitigate such risks must integrate a game-theoretic model to effectively capture these interactions \cite{manshaei2013game,zhu2018game,rass2020cyber}. Illustrated in Figure \ref{dynamics}, the dynamics between the defender and the attacker need  a dynamic game, as depicted in Figure \ref{game}, denoted by a sequence of games $G_1, G_2, \cdots, G_6$. Each game Gt encapsulates the local interaction at stage $t$. A fundamental game comprises three key components: the set of players, their action sets, and their payoff functions. The sequences of games can capture different stages of resilience \cite{zhu2018multi}. For example, proactive resilience mechanisms can be captured by one game, while responsive resilience mechanisms can be captured by another. Retrospective resilience can also be captured by a separate game. Each mechanism involves different types of players and different actions. For example, responsive resilience involves the sequential tactics and procedures of the attacker aimed at advancing toward the target. Proactive resilience involves the attacker's strategy to exploit vulnerabilities on the attack surface, while retrospective resilience involves analyzing the entirety of the attacker's operation. Hence, the defender's tactics and strategies will vary according to the stage involved.

\begin{figure}[ht]
\begin{center}
    \includegraphics[scale=0.7]{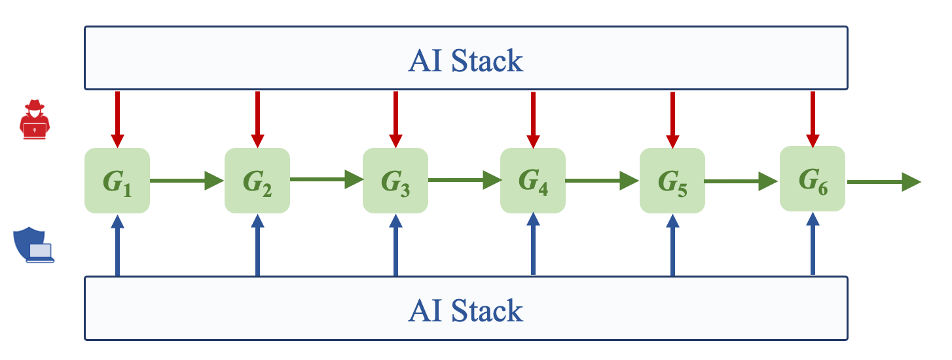}
    \caption{The interactions between the attacker and the defender utilizing resilience mechanisms can be depicted through a sequence of games at various stages. The actions taken by both parties shape the progression of the games in subsequent steps. Resilience mechanisms can be augmented by an AI stack that facilitates the analysis of adversarial behaviors and strategically adjusts its own policies throughout the stages of the games.}
\end{center}
\end{figure}\label{game}

Each described game represents the outcome of the interplay between threats posed by adversaries and defense mechanisms employed by defenders. The equilibrium of these games, often described using Nash equilibrium, which is a point of strategies  where no players have incentives to deviate from unilaterally.  It serves as a primary metric to gauge the consequence of these interactions, as depicted in Figure \ref{metric}. This equilibrium offers insights into the predicted outcome $Y_t$. Here, the risk  $r$ can be thoroughly assessed, not just on an average basis across all threat events, but also considering the strategic maneuvers of the attacker, who intelligently selects the most impactful vulnerability and associated event $E_t$ at time $t$ to achieve their goal $G_A(E_t, R_t)$. Here, $G_A$ is the attacker’s reward function, which takes as input the threat event chosen at the attacker at time $t$ and the defender’s resilience action $R_t$ at time $t$. It yields a reward that indicates the proximity toward the goal.  

Similarly, the defender's objective is to select a resilience mechanism $R_t$ at time $t$ to achieve their own goal $G_R(E_t, R_t)$. Here $G_R$ is the defender’s resilience reward function, which is  influenced not only by the defense action $R_t$ but also by the attacker's action $E_t$. Both players aim to maximize their reward by choosing respective policies. The equilibrium outcome, denoted by $E_t^*$ and $R_t^*$, is a point where no players can benefit from deviating from their actions. As suggested in Figure \ref{risk}, this outcome will further result in the equilibrium state $X_t^*$ and the associated consequence $Y^*_t$. This consequence $Y^*_t$ serves as a key metric for risk evaluation within this paradigm. We call this risk as strategic cyber risk. Its evaluation considers the strategic behaviors of both attackers and defenders, offering meaningful predictions when agents behave rationally and engage in sufficient reasoning over time. Strategic risk is particularly valuable when event probabilities are unavailable, a common occurrence in cybersecurity scenarios. Obtaining data to assess how frequently attackers exploit the attack surface can be challenging, and often the precise attack surface is not fully understood. Consequently, strategic risk offers an alternative approach. Its evaluation needs an attacker model that quantifies attacker actions and motivations. Risk assessment involves considering potential outcomes given both the rational behaviors of attackers and the network's best effort defense. While assuming rationality and understanding exact incentives may be difficult, these assumptions provide a foundational framework. Nevertheless, this framework can be enhanced through the incorporation of bounded rationality models \cite{chen2019interdependent,hu2023game,hu2023detection,thakoor2020exploiting,rass2020bounded} and learning methodologies \cite{zhu2012hybrid,zhu2011distributed,zhu2022introduction}, allowing for continual refinement and improvement.

The games depicted in Figure \ref{game} are interleaved, with the outcome of a game at stage t determining the subsequent game. The dynamic game is characterized by a transition kernel that outlines the transition between games. Dynamic games can manifest in various forms contingent upon the players' information structure, the dynamics' structure, and the uncertainties in the games. For instance, in \cite{zhu2009dynamic,zhu2010network}, a dynamic stochastic game has been employed to configure intrusion detection systems across different system states. This intrusion detection game involves a defender configuring rules to detect the attacker, who, in turn, aims to evade detection.

Bayesian games are utilized to capture situations where players have incomplete information about the game. One source of uncertainty arises from the payoff of the game. The players do not know the exact payoff of the game but rather the probabilities of the outcomes. Sometimes, the players may not know the exact game they are playing, including the actions they can take and the other players' actions. Harsanyi has provided methods to understand such games with uncertainties by adopting a Bayesian approach \cite{harsanyi1967games}. The unknowns of a player are characterized by the notion of type, representing the private information or characteristics that affect one player's payoffs or strategies in the game. Each player's type is drawn from a probability distribution known to the player but not necessarily to others. It is often assumed that the players share beliefs about the distribution of types or the uncertainty in the game. This assumption is known as the common prior, which serves as a basis for reasoning and decision-making in the game.

Another approach is based on Aumann’s knowledge models \cite{aumann1995epistemic,aumann1976extend,aumann2002incomplete,aumann1974subjectivity}. It makes use of epistemic logic to deal with reasoning about knowledge and belief. Epistemic logic provides formal tools for representing and reasoning about agents' knowledge and beliefs, as well as the interactions between them. One key notion of Aumann’s approach is the notion of an information partition. This partition divides the set of possible states of the world into distinct subsets, each corresponding to a particular piece of information available to players. The partition captures the structure of players' information and determines how it influences their beliefs and strategies. In Aumann’s approach, common knowledge is defined as the information that is not only known by an individual agent but is also known to be known by all agents, and so on iteratively. It connects with the common prior assumption in Harsanyi’s model. It has been shown that Harsanyi’s Bayesian games and Aumann’s epistemic approach are equivalent representations of games of incomplete information.

In cases where one agent holds more information than the other, asymmetric information games arise, often used to model cyber deception contexts. For example, they have been used in honeypot deployment and design \cite{horak2017manipulating}, where the attacker is assumed to have incomplete information about the honeypot's location, but the defender has complete information about it. On the other hand, in the application of designing insider threat mitigation programs, the attacker has complete knowledge of whether they are an insider or not, but the defender has incomplete information about who the insiders are \cite{huang2021duplicity,huang2020dynamic}. Asymmetric information games provide a modeling framework to understand the information advantage of the players and find ways to either reduce this advantage gap or leverage it to mitigate cyber risks.

One special class of asymmetric information games is signaling games \cite{casey2016compliance,zhang2018hypothesis,mohebbi2015trust,pawlick2017strategic,hu2022evasion}, wherein a sender observes the true state of the world, crafts a message, and sends it to a receiver to elicit a response aligned with the sender's intentions. The receiver, upon observing the message, must estimate the true state of the world based on the sender's strategies. Lack of consistency arises if the receiver's belief about the true state of the world is not supported by the messages sent by the sender as per the sender’s strategy. Hence, the equilibrium concept of this game involves not only strategies that do not allow unilateral deviation but also the sender's associated belief with these strategies. This equilibrium concept is called Bayesian Nash equilibrium. This class of games has contributed to understanding information manipulation, man-in-the-middle attacks, and misinformation over networks. 

For example, in \cite{xu2015cyber,pawlick2015flip,zhu2020cross}, a man-in-the-middle attack between an industrial control system and a cloud is described. An industrial control system sends a computation task to the cloud for heavy computations, including image processing, control computation, or state estimation. The cloud then sends back the results to the control systems, which will utilize the results to actuate the system. The cloud-enabled control system leverages the computational power at the edge to reduce the computation burden on the system itself, which has limited real-time computation power due to hardware constraints. However, the attacker can manipulate the result and send a false result to mislead the control system into a failure mode of operation. One approach to creating resilience in case cryptography fails is to implement a checking mechanism to verify whether the sent message is coherent. Utilizing a belief system can aid in this verification process. The signaling game framework naturally captures the fact that the attacker has complete information and sends a message to the control system, which then must decide whether to utilize the result or discard it. Developing a strategy to screen the inputs provides a layer of resilience to the system against such attacks.

Another example is the zero-trust access control problem \cite{ge2022mufaza,ge2023zero,li2023decision}. In this scenario, the attacker knows their type, but the network does not. The network must decide whether to grant or revoke access based on the behavior or footprint of the user. This situation corresponds to a dynamic signaling game where the attacker has complete information about their type, but the network does not. A game-theoretic approach offers a strategy for authentication based on the dynamic formation of a belief process, which relies on continuous monitoring. This plays a critical role in providing zero-trust security to modern networks, where access is determined through continuous monitoring.

Game theory not only captures scenarios of interactions and provides a way to determine the optimal strategy to defend against threats but also facilitates the design and deployment of overarching resilience mechanisms. In this regard, mechanism design theory \cite{myerson1989mechanism,borgers2015introduction} becomes essential as it enables us to shape the equilibrium or outcome of the game to a desirable one through the design of the game's rules. One such framework is Stackelberg games \cite{liu2022stackelberg,maharjan2013dependable,pawlick2016stackelberg}. Stackelberg games typically consist of a leader and several followers. The leader, acting as a designer, anticipates the outcomes of the game played by the followers. The followers interact with each other based on the input from the leader. The leader then designs an appropriate input to the game to shift the outcome to achieve the goal of the design.  Such designs have been utilized in insurance schemes for computer networks \cite{zhang2017bi,zhang2019mathtt,zhang2021optimal}. The insurer must design a scheme that enhances the network's security ecosystem and resilience. To achieve this, the insurer needs to assess risks by considering the interactions between attackers and defenders. The equilibrium of the game framework indicates the risks associated with the network under the designed insurance mechanism. The objective of the design is to minimize the insurer's costs while aligning with the cyber risks of the insuree. This framework has been extended to dynamic and network settings.

Bayesian mechanism design theory plays another pivotal role in shaping game outcomes to achieve cyber resilience. It has been employed to understand and influence player behaviors within a novel game-theoretic framework known as the duplicity game \cite{huang2021duplicity}, aimed at designing effective deception mechanisms. This framework, comprising a generator, an incentive modulator, and a trust manipulator, collectively referred to as the GMM mechanism, is formulated as a mathematical programming problem. Its objectives include computing the optimal GMM mechanism, quantifying the upper limit of enforceable security policies, and characterizing conditions for user identifiability and manageability in terms of cyber attribution and user management.

Dynamic game modeling provides a foundational framework for understanding the cyber environment, as illustrated in Figure \ref{risk}. Various games capture distinct types of interactions within this environment. The dynamic game framework needs an investigation with a forward-looking perspective, meaning actions taken now must consider not only the current stage but also subsequent stages of the game.

Equilibrium analysis of dynamic games within this forward-looking context often employs dynamic programming principles or moving-horizon schemes, which are fundamental control-theoretic techniques utilized to tackle long-term problems. The control framework introduced in \ref{riskmetric} is extended to a dynamic game framework with a more intricate modeling of adversarial behaviors, thereby creating a multi-player control problem. Such analysis yields a risk trajectory of the network environment in the face of adversaries. This analysis facilitates the measurement of network risks based on the prescribed attack model. The sequence of stage-by-stage risks corresponds to the cyber resilience metric depicted in Figure \ref{metric}. In essence, at each stage, a game can assess risks by predicting the consequences of interactions between threats and defenses.

The strategies associated with dynamic games offer a means to manage the evaluated risks under given attack models. If the assessed risks meet the performance criterion, the game strategies can inform the design of cyber resilience mechanisms at the operational and tactical levels, informing network actions how to act to maintain an acceptable level of risk. However, if the assessed risks do not meet the performance criterion, mechanism design theory enables a further step towards planning and designing resilience mechanisms to ensure the game arrives at a desirable set of outcomes with acceptable risks. These designed games correspond to the design of networks, rules, and policies, thereby informing the strategic-level planning and deployment of resilience mechanisms.

\subsection{Learning Foundations}

Despite advances in game-theoretic modeling and related computational and mechanism tools, challenges persist in utilizing game theory for cyber resilience. One key challenge is the imprecise modeling or capture of threat models. Attack behaviors can surpass prescribed models, leading to inaccurate risk assessments and potentially misleading designs due to misalignment between actual threats and prescribed ones. This issue is inevitable as networks face zero-day threats. Even if models initially capture exact attack behavior, the rapidly evolving threat landscape poses challenges for modeling to keep pace. 

In addition to the epistemic uncertainties associated with threat models, aleatoric uncertainties related to game parameters such as payoffs, action sets, common priors, and dynamics remain challenging to precisely determine. Therefore, learning emerges as a crucial approach to enhance the game and control framework, enabling data-driven, autonomous, and agile defense strategies for resilience. There is a pressing need to develop an AI stack, as depicted in Figure X, which utilizes dynamic game models to facilitate advanced reasoning about attackers. This stack should also incorporate computational and control techniques to compute lookahead policies, allowing for online observations to adapt resilience policies dynamically and defend against attackers.

Recent advances in learning and AI  have introduced numerous models and approaches to cybersecurity. For instance, leveraging large language models can expedite information processing and learning, facilitating rapid parsing of network log and traffic data for improved decision-making. Deep neural networks have also notably enhanced intrusion detection. Among learning paradigms, reinforcement learning emerges as a vital element for enabling autonomous and data-driven resilience mechanisms. Figure \ref{learning} illustrates the feedback structure of reinforcement learning.   In this paradigm, the network system is subject to continuous monitoring, ensuring a continuous watch over its operations and interactions. This monitoring is followed by a sophisticated reasoning framework, designed to analyze incoming data and promptly identify any deviations or anomalies within the network's behavior. Leveraging this analytical capability, the framework dynamically updates network policies in real-time, tailoring responses to specific network events as they unfold. This adaptive approach not only enables fast detection of potential threats but also facilitates proactive risk mitigation strategies.  This paradigm aligns well with the control framework detailed in Section X, where a feedback structure is integral for enabling dynamic responses and automated mechanisms. 

\begin{figure}[ht]
\begin{center}
    \includegraphics[scale=0.7]{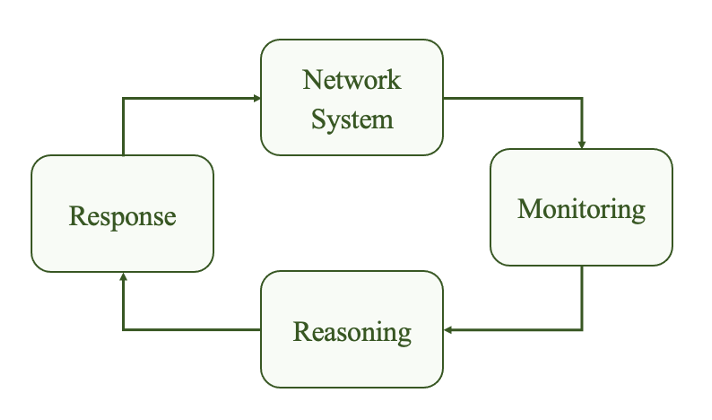}
    \caption{Reinforcement learning offers a paradigm for achieving adaptive resilience mechanisms. The network system undergoes continuous monitoring, with a reasoning framework employed to update network policies in response to events within the network. This adaptation aims to mitigate risks to the network system.}
\end{center}
\end{figure}\label{learning}

In \cite{tao23ztd}, meta-learning reinforcement has been used to create network defense adaptable to various attack scenarios. Meta-learning is a machine learning technique that focuses on learning how to learn. It involves training models on a variety of tasks or datasets, with the goal of enabling the model to quickly adapt to new tasks or scenarios with minimal additional training. In the context of defense strategies, meta-learning is used to enhance the adaptability and generalization capabilities of the system in response to evolving and diverse attack scenarios.
A meta policy is first learned through meta-learning. This meta policy encapsulates a high-level strategy for making trust-related decisions based on partial observations of the system and potential attack scenarios. Along with the meta policy, an adaptation mapping is learned using meta-learning techniques. This mapping allows the system to quickly adjust the meta policy based on new information or unseen attack scenarios, enabling rapid adaptation to changing threat landscapes. By leveraging meta-learning, the defense system can efficiently adapt to new and unknown attack scenarios using insights gained from previous training data. This adaptive capability reduces the need for extensive retraining or manual intervention when faced with novel threats.

In \cite{li2024conjectural}, Conjectural Online Learning (COL) has been proposed as a learning scheme designed for generic Asymmetric Information Stochastic Games (AISGs) where decision-making entities have limited or asymmetric information. COL utilizes a forecaster-actor-critic (FAC) architecture to update strategies in an online manner by incorporating first-order beliefs and subjective forecasts of opponent strategies. In the context of security, COL can be applied to enhance cyber resilience in IT infrastructures. For example, in defending against Advanced Persistent Threats (APTs), COL can help in formulating strategies for the defender to protect the infrastructure and respond to intrusions effectively. By utilizing Bayesian learning and online adaptability, COL can enable defenders to adjust their strategies based on evolving threats and nonstationary attack patterns, leading to improved resilience against cyber threats.

In \cite{li2024symbiotic}, the interconnections among foundation models (FMs), game models, and learning are explored. FMs serve as essential components for constructing customized machine learning models tailored to diverse cybersecurity applications. Game-theoretic models provide a foundational framework for comprehensively modeling adversarial interactions, facilitating the encapsulation of adversarial knowledge and domain-specific insights. The synergy between GMs and FMs can yield proactive and automated cyber defense mechanisms. These mechanisms not only aim to secure networks against attacks but also endeavor to enhance resilience against well-planned operations. It has been highlighted that one promising direction is the multi-agent neurosymbolic conjectural learning (MANSCOL) approach, which empowers defenders to predict adversarial behaviors, design adaptive defensive deception tactics, and synthesize knowledge for operational-level synthesis and adaptation. FMs assume a pivotal role in various functions within MANSCOL, encompassing reinforcement learning, knowledge assimilation, formation of conjectures, and contextual representation.

\subsection{Confluences of Control, Games, and Learning Methods}

Control, game, and learning methods each play pivotal roles in bolstering cyber resilience. Control-theoretic methods offer a dynamic systems perspective, framing cyber resilience as a continual process where defense mechanisms must operate over time across multiple stages to mitigate risks stemming from the attack surface. Conversely, game-theoretic methods distinguish cyber resilience from resilience to natural disasters by modeling adversarial behaviors of attackers. These methods facilitate the assessment of strategic risks faced by defenders under the strategic behaviors of attackers. Additionally, leveraging mechanism design and computation tools from game theory and control theory enables the dynamic design of cyber resilience mechanisms. Their confluences are central to the theoretic foundations of cyber resilience as illustrated in Figure \ref{confluence}.

\begin{figure}[ht]
\begin{center}
    \includegraphics[scale=0.7]{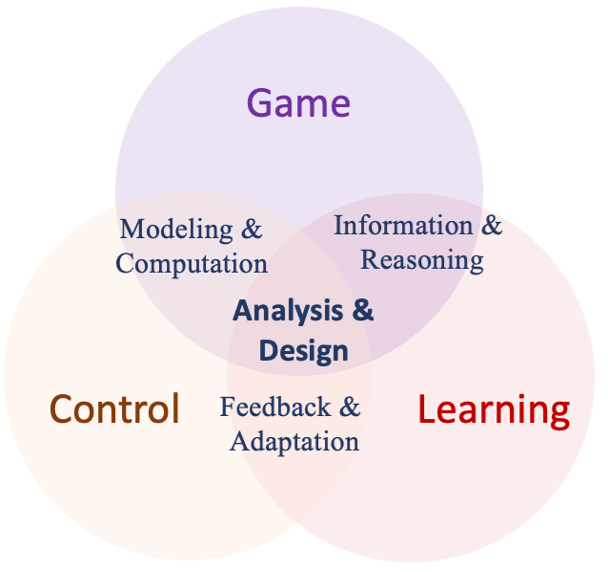}
    \caption{The convergence of game, control, and learning methods is essential. Game and control methods establish the groundwork for modeling and computations in cyber risk and its mitigation. Control and learning methods are pivotal in the development of feedback-driven and adaptive resilience mechanisms. Game and learning methods offer data-driven strategic reasoning mechanisms utilizing both symbolic and data information. Their convergence enables the analysis and design capabilities crucial for cyber resilience.}
\end{center}
\end{figure}\label{confluence}

Furthermore, learning methods facilitate the integration of data and symbolic information from models, enabling the creation of an implementable AI stack. This stack enables analysis of adversarial behaviors, real-time response of defenses, and adaptive design and planning of mechanisms. By integrating these three methods, machine intelligence can be harnessed to address threats in an automated, adaptive, and agile manner, fostering new innovations in the realm of AI and cyber resilience solutions.

Recent advances in predictive learning with adversarial thinking  exemplify the synthesis of control, game, and learning methods. By effectively learning the adversarial model, valuable insights into adversaries' tactics are gained, enabling the development of robust response strategies. Predictive control methods, such as roll-out and forward-looking strategies, complement predictive learning by facilitating rapid adaptation and decision-making in dynamic environments.

This advance is closely tied to the concept of conjectural learning \cite{li2024conjectural,tao23cola}, wherein agents hypothesize about adversaries' behavior based on observed actions and outcomes. Incorporating conjectural learning enhances the ability to anticipate and respond to adversarial maneuvers, thereby improving overall resilience and adaptive capacity.

Additionally, meta-learning, which involves learning to learn, plays a significant role in this synthesis \cite{pan-tao23meta-sg,tao23ztd}. By leveraging meta-learning techniques, adaptability and optimization of response strategies based on evolving threats and changing environmental conditions are enhanced. Meta-learning facilitates the extraction of meaningful patterns and insights from past experiences, enabling more effective decision-making and strategy formulation.

An essential aspect of this synthesis involves coordinating and managing tradeoffs among different stages of the resilience process. Each stage, from proactive preparation to reactive response and retrospective analysis, presents its own set of challenges and considerations. Effective coordination between these stages is crucial for ensuring a cohesive and comprehensive approach to resilience. By carefully balancing tradeoffs and leveraging insights from each stage, resource allocation can be optimized, actions prioritized, and overall resilience effectiveness maximized.

\subsection{Cyber deception as a case study}

Modern advances in cyber deception for cyber resilience is an example of the integration of control, games, and learning within the cybersecurity domain. By deploying deceptive techniques, organizations can manipulate the behavior of adversaries, influence the dynamics of the cyber environment, and enhance their resilience against threats.

Control mechanisms play a crucial role in cyber deception by governing the deployment and management of deceptive techniques and assets. Through control mechanisms, organizations can strategically design and implement deceptive environments to manipulate adversary behavior and mitigate threats \cite{fang2022fundamental,fang2021fundamental,fang2021fundamental,peng2019game}. Control aspects ensure that deceptive measures are deployed effectively and efficiently, maximizing their impact on adversaries while minimizing the risk to legitimate systems and data. In the context of resilience, control enables defenders to adapt and optimize their deception strategies in response to evolving threats and changing environmental conditions, thereby enhancing overall cyber resilience.

Game theory provides a strategic framework for understanding and influencing the interactions between defenders and adversaries in the cyber domain \cite{pawlick2019game,pawlick2021game,pawlick2019game}. In the context of cyber deception, game theory principles guide the design and implementation of deceptive strategies to outmaneuver adversaries and increase the defender's advantage. By presenting adversaries with strategic dilemmas and uncertainties, defenders can exploit decision-making vulnerabilities and manipulate adversary behavior to their advantage. Game theory also facilitates the analysis of adversary strategies and the identification of optimal deception strategies for enhancing resilience against cyber threats.

Learning mechanisms enable defenders to adapt and improve their deception strategies based on insights gained from past experiences and interactions with adversaries \cite{Tao_blackwell,tao22confluence,huang2022reinforcement}. Through machine learning algorithms and data analytics, defenders can analyze adversary behavior, identify patterns, and refine deception tactics over time. Learning enables defenders to anticipate evolving threats, proactively adjust deception strategies, and continuously enhance resilience against cyber attacks. By incorporating learning mechanisms into cyber deception and resilience strategies, organizations can improve their ability to detect, deter, and mitigate threats in dynamic and evolving cyber environments.

The integration of control, game theory, and learning principles enhances the effectiveness of cyber deception and resilience strategies. Control mechanisms govern the deployment and management of deceptive assets, ensuring strategic placement and utilization. Game theory principles guide the design of deceptive environments to exploit adversary decision-making vulnerabilities. Learning mechanisms enable defenders to adapt and optimize deception strategies based on insights gained from past interactions with adversaries. Together, these integrated approaches enable organizations to enhance their cyber deception capabilities and resilience against evolving cyber threats.

\section{Future Directions}
Cyber resilience encompasses a diverse array of challenges. This domain transcends mere technicalities, extending to the intersections of cyber, physical, and human components within systems. Many of these challenges are interdisciplinary, necessitating a systems-scientific approach for resolution. Furthermore, within the realm of tools, there exists a wealth of opportunities at the intersection of game theory, control theory, and machine learning. Numerous methods can draw inspiration from emerging applications and be developed to tackle a wide range of issues.

On the methodological front, non-equilibrium learning, large language models, and foundation models are emerging as promising tools to enhance the intelligence associated with cyber resilience mechanisms. Nonequilibrium learning is a novel approach in machine learning and artificial intelligence that diverges from traditional equilibrium-based models \cite{pan-tao22noneq,pan-tao23delay}. Unlike equilibrium-based approaches that assume stable states and predictable outcomes, nonequilibrium learning focuses on dynamic, non-stationary environments where conditions are constantly evolving. This approach is particularly relevant in cybersecurity, where adversaries continually adapt their tactics to circumvent defenses. Nonequilibrium learning algorithms are designed to adapt to changing conditions in real-time, enabling defenders to detect and respond to emerging threats more effectively. By leveraging nonequilibrium learning, organizations can enhance their ability to anticipate and mitigate cyber attacks in dynamic and unpredictable environments.

Large language models (LLMs), such as Generative Pre-trained Transformer (GPT) models, have revolutionized natural language processing. These models, trained on vast amounts of text data, possess advanced capabilities in generating human-like text and understanding contextual nuances. In cybersecurity, LLMs are increasingly being applied to analyze and understand adversarial behavior in strategic games. By training LLMs on extensive datasets of cyber threat intelligence and historical attack patterns, researchers can develop models capable of predicting adversary strategies and identifying potential vulnerabilities. These LLM-based approaches provide valuable insights into the strategic dynamics of cyber conflicts, enabling defenders to anticipate and counteract adversarial actions more effectively.

Foundation models with strategic learning represent a paradigm shift in the development of AI and machine learning algorithms for cybersecurity \cite{li2024symbiotic}. Unlike traditional models that focus solely on optimizing specific tasks or objectives, foundation models with strategic learning incorporate strategic decision-making capabilities inspired by game theory and multi-agent systems. These models are designed to analyze and predict the behavior of multiple actors in complex cyber environments, taking into account their strategic interactions and incentives. By integrating strategic learning into foundation models, researchers can develop more robust and adaptive cybersecurity solutions capable of addressing the strategic challenges posed by sophisticated adversaries. These models enable defenders to anticipate adversary tactics, identify strategic vulnerabilities, and devise effective countermeasures to enhance cyber resilience.

Besides the advances in methodologies, there are also increasingly challenging cyber risk issues that we need to address. One key challenge associated with cyber resilience is the interdependencies among the risks within the subnetworks. Developing resilience mechanisms for the entire network itself is not a scalable solution. Instead, creating decentralized mechanisms for individuals is more plausible. However, in this case, the resilience mechanisms need to coordinate with each other due to the interdependencies, while simultaneously controlling and managing the cyber risks within their own network. The cyber risk of the entire network under this cyber resilience architecture is called holonic risk. Holons are entities that constitute the whole system, but each is capable of autonomous functions \cite{van1998reference}. Holonic risks refer to the risk of the whole network, in which each constituent subnetwork autonomously manages its risk. This type of risk is becoming more common as the network grows larger and more heterogeneous, and each subnetwork is owned by different entities who have the right to control it. One way to deal with such issues relies on game-theoretic, control-theoretic, and learning tools. The interdependencies among the subnetworks can be viewed as a game in which individuals respond to each other. In this case, a subnetwork not only needs to dynamically respond to threats but also to the behaviors of other subnetworks, bringing a new layer of complexity into the cyber resilient framework. Similarly, the associated learning involves not only learning the behaviors of adversaries but also the behaviors of other networks through communications and cooperation.

The consequences of cyber risks in critical infrastructures and industrial networks often extend to risks within physical systems. Consequently, the risk is not solely confined to the cyber domain but incorporates both cyber and physical domains, making it a cyber-physical challenge. Thus, associated resilience mechanisms must not only address cyber threats but also consider the implications for physical systems. Therefore, the design of resilience mechanisms needs to be physics-informed, i.e., integrating insights from physics to ensure compliance with operational level constraints. Additionally, it is essential to develop physical-level resilient mechanisms, such as controllers, to mitigate impacts on physical systems. Consequently, cyber resilience must be jointly designed with physical resilience, emphasizing a co-design paradigm for networks \cite{zhu2020cross,huang2021cross,xu2018cross,yuan2013resilient,huang2020dynamic}. This approach is crucial and has been applied in understanding cyber-physical resilience across various domains, including nuclear systems \cite{zhao2020finite,zhao2019game,smidts2022cyber}, power systems \cite{zhu2011robust,zhu2015game,chen2016game,zhu2019multilayer,zhu2011secure}, industrial control systems \cite{collier2016security,chen2021dynamic,zhao2024integrated,xu2020secure,xu2015secure,farooq2018adaptive}, multi-agent systems \cite{zhao2022multi,nugraha2020dynamic,rieger2012agent,chen2019control,nugraha2020dynamic,nugraha2021rolling}, and interdependent infrastructures \cite{chen2019game,chen2019dynamic,huang2018distributed,
huang2020dynamic,rinaldi2001identifying}.

Incorporating emerging methods into cybersecurity research and practice holds great promise for advancing our understanding of cyber threats and improving our ability to defend against them. By leveraging nonequilibrium learning, large language models, and foundation models with strategic learning, organizations can stay ahead of evolving threats and improve their cyber resilience in an increasingly complex and dynamic threat landscape.

\section{Conclusions}
Dealing with cyber risk in today's world is increasingly challenging. Mitigating cyber risk should not only involve cyber security defense mechanisms but also cyber resilience mechanisms, which complement each other to strengthen the cyber risk posture of the network. Understanding cyber risk and the associated resilience mechanisms requires the development of fundamental principles, metrics, and tools to analyze and design networks to achieve resilience goals. This chapter introduces game theory, control theory, and learning theory as the foundation for cyber resilience. Game theory has been shown to be suitable for modeling strategic interactions. Control theory has been pivotal in dealing with dynamics and designing dynamic defenses. Learning theory plays an important role in bridging theory and practice through autonomous adaptation and dealing with uncertainties. 
The confluence of game theory, control theory, and learning theory emerges as a cornerstone for cyber resilience, enabling organizations to improve their cyber resilience posture. As cyber threats continue to escalate in complexity and frequency, the interdisciplinary framework presented in this chapter serves as a roadmap for organizations seeking to enhance their cyber resilience capabilities and navigate the intricate dynamics of modern cybersecurity challenges.

\bibliographystyle{abbrv}

\bibliography{reference.bib} 
\end{document}